\documentclass[11pt]{article}
 \csname
@addtoreset\endcsname{equation}{section}

\def\a{\alpha}
\def\b{\beta}
\def\c{\gamma}
\def\C{\Gamma}
\def\d{\delta}
\def\D{\Delta}
\def\e{\epsilon}
\def\ve{\varepsilon}
\def\f{\phi}

\def\k{\kappa}
\def\l{\lambda}
\def\L{\Lambda}
\def\m{\mu}
\def\n{\nu}
\def\r{\rho}
\def\s{\sigma}

\def\x{\xi}

\def\o{\omega}
\def\pb{{\bar\psi}}
\def\eb{{\bar\varepsilon}}
\def\lb{{\bar\lambda}}

\def\cO{{\cal O}}

\def\vx{{\vec x}}

\def\cO{{\cal O}}

\def\ra{\rightarrow}

\def\qq{\quad\quad}
\def\qqq{\quad\quad\quad}

\def\del{\partial}
\newcommand{\la}[1]{\label{#1}}
\let\bm=\bibitem
\def\nn{\nonumber}
\newcommand{\eq}[1]{(\ref{#1})}

\newcommand{\w}[1]{\\[0.#1cm]}
\def\eqs#1#2{(\ref{#1}-\ref{#2})} 

\def\be{\begin{equation}}
\def\ee{\end{equation}}
\def\bea{\begin{eqnarray}}
\def\eea{\end{eqnarray}}
\def\ba{\begin{array}}
\def\ea{\end{array}}

\def\se{\;\;=\;\;}

\def\mx#1#2#3#4{\left#1\begin{array}{#2} #3 \end{array}\right#4}

\def\ft#1#2{{\textstyle{{\scriptstyle #1}
\over {\scriptstyle #2}}}}

\def\Hat#1{\widehat{#1}}

\def\ed{\end{document}}

\newcommand{\hoch}[1]{$\, ^{#1}$}

\newcommand{\tamphys}{\it\small Center for Theoretical Physics, Texas
A\&M University, College Station, TX 77843, USA}

\newcommand{\groningen}{\it\small Institute for Theoretical Physics,
Nijenborgh 4, 9747 AG Groningen,The Netherlands}

\newcommand{\saitama}{\it\small Physics Department, Faculty of Science,
Saitama University, Urawa, Saitama 338-8570, Japan}

\newcommand{\auth}{\large N.S. Deger \hoch{1}, A. Kaya \hoch{1}, E.
Sezgin \hoch{1}, P. Sundell \hoch{2} and Y. Tanii \hoch{3}}

\begin{document}

\hfill{CTP-TAMU-38/99}

\hfill{UG/22-00}

\hfill{STUPP-00-161}

\hfill{hep-th/0012139}


\vspace{20pt}

\begin{center}

{\Large \bf (2,0) Chern-Simons Supergravity Plus Matter \w2
{\Large \bf Near the Boundary of $AdS_3$}}

\vspace{30pt}


\auth

\vspace{15pt}

\begin{itemize}

\item[$^1$] \tamphys

\item[$^2$] \groningen

\item[$^3$] \saitama

\end{itemize}

\vspace{30pt}

{\bf Abstract}

\end{center}

We examine the boundary behaviour of the gauged $N=(2,0)$
supergravity in $D=3$ coupled to an arbitrary number of scalar
supermultiplets which parametrize a K\"ahler manifold. In addition
to the gravitational coupling constant, the model depends on two
parameters, namely the cosmological constant and the size of the
K\"ahler manifold. It is shown that regular and irregular boundary
conditions can be imposed on the matter fields depending on the
size of the sigma model manifold. It is also shown that the super
AdS transformations in the bulk produce the transformations of
the $N=(2,0)$ conformal supergravity and scalar multiplets on the
boundary, containing fields with nonvanishing Weyl weights
determined by the ratio of the sigma model and the gravitational
coupling constants. Various types of (2,0) superconformal
multiplets are found on the boundary and in one case the
superconformal symmetry is shown to be realized in an
unconventional way.

{\vfill\leftline{}\vfill


\pagebreak

\setcounter{page}{1}


\section{Introduction}


In probing various aspects of the remarkable connections between
anti de Sitter and conformal supergravity theories, the
$AdS_3/CFT_2$ correspondence in particular provides a relatively
more manageable case to study. At the same time, some novel
features arise due to the fact that AdS supergravity in $D=3$ is
essentially non-dynamical. Nonetheless, $AdS_3$ supergravity plays
a significant role in the description of the matter fields to
which it couples. As a step towards a detailed study of the
amplitudes, anomalies and other significant properties of this
type of theories, it is useful to determine precisely the
behaviour of the AdS supersymmetry transformations at the
boundary. This problem has been examined for pure $N\le 2$ $AdS_3$
supergravity in \cite{t1}, pure $N=4$ $AdS_3$ supergravity in
\cite{weyl}
\footnote{Various aspects of the AdS/CFT correspondence for pure
supergravities in $D=3$, in particular the asymptotic symmetries,
have been studied in \cite{hen1,hen2}.},
the maximal $AdS_7$ supergravity in \cite{t2}, the maximal $F(4)$
$AdS_6$ supergravity in \cite{y} and the minimal $AdS_5$
supergravity in \cite{4d}, where it was shown that the correct
transformations rules of the boundary conformal supergravities
indeed follow from a careful study of the bulk super AdS
transformations. However, a similar analysis does not seem to have
been carried out so far for matter coupled AdS supergravities
which should shed further light on the AdS/CFT duality questions
in the context of M-theory in backgrounds with less than maximal
supersymmetry. Our aim in this paper is to fill this gap.

AdS supergravities are based on AdS superalgebras. Given the fact
that the AdS group in $2+1$ dimensions is a product of two factors
as $SO(2,2) = SO(2,1)_L \times SO(2,1)_R$, the super AdS group
itself has the factored form $G_L \times G_R$. It turns out that
there are many choices for $G_{L,R}\,$, the most typical case
being $OSp(2,p) \times OSp(2,q\,)$, where $p$ and $q$ are not
necessarily equal. The supergravity theories based on these
algebras will be referred to as the $N=(p,q)$ $AdS_3$
supergravities. They have been constructed as Chern-Simons gauged
theories long ago by Achucarro and Townsend \cite{at}. However,
very little is known so far about their matter couplings. In fact,
the only cases studied until now seem to be the $N=(2,0)$ $AdS_3$
supergravity coupled to an arbitrary number of scalar
supermultiplets \cite{it,mc} and $N=(16,0)$ $AdS_3$ supergravity
with an exceptional sigma model sector \cite{nic}. The models
constructed in \cite{it} and \cite{mc} are significantly different
from each other, stemming from the fact that the scalar fields are
neutral under the U(1) R-symmetry group in the model of \cite{it},
but charged in the model of \cite{mc}. The Izquierdo-Townsend
model has only one free parameter, namely the cosmological
constant, in addition to the gravitational constant, unlike the
model of \cite{mc}, where there is the additional parameter that
measures the size of the sigma model manifold. In fact, the U(1)
charge carried by the scalar fields is related to this size, and
as we will show in this paper, the limit in which the U(1) charge
vanishes implies a flat sigma model manifold, and the models of
\cite{it} and \cite{mc} do indeed agree in that case.

The model studied here is expected to arise from a
compactification of $M$-theory. The much studied compactification
of Type IIB theory on $AdS_3 \times S^3 \times K$, where $K$ is
essentially $T^4$ or $K3$, gives rise to $N=(4,4)$ or $N=(4,0)$
$AdS_3$ supergravities coupled to matter. The spectra of these
theories are known \cite{dkss,db} but not their actions so far.
Whether our $N=(2,0)$ model arises as a consistent truncation of
such theories remains to be seen.

The super AdS transformations in the bulk theory studied here are
shown to produce the transformations of $N=(2,0)$ conformal
supergravity coupled to scalar multiplets with nonvanishing Weyl
weight determined by the ratio of the K\"ahler sigma model
manifold and the gravitational coupling constant. In doing so, the
so called regular and irregular boundary conditions are utilized
\cite{bkl,kw}. These choices of boundary conditions result in the
phenomenon in which scalar fields in AdS space of sufficiently
negative mass-squared can be associated with CFT operators of two
possible dimensions. An example of this has been discussed in
\cite{kw} in the context of $AdS_5 \times T^{1,1}$
compactification of Type IIB string theory. Here, we provide
another example of this phenomenon and show explicitly the
resulting CFT supergravity plus matter symmetry transformations.
In doing so, we find an interesting conformal supermultiplet
structure that involves a submultiplet of fields that transform
into each other. In this novel multiplet the superconformal
symmetry is realized in an unconventional fashion.

The $(2,0)$ model is described in the next section. The relation
between the models of \cite{it} and \cite{mc} is described in
Section 3. The boundary conditions and the linearized field
equations are given in Section 4, and the bosonic and fermionic
symmetries of the boundary CFT are obtained in Section 5.
Concluding remarks are contained in Section 6.


\section{The Matter Coupled N=(2,0)\ AdS$_3$ Supergravity
\label{sectwo} }


The $N=(2,0)\ AdS_3$ supergravity multiplet consists of a graviton
$e_\m{}^a$, two Majorana gravitini $ \psi_\m$ (with the $SO(2)$
spinor index suppressed) and an $SO(2)$ gauge field $A_\m$. The $n$
copies of the $N=(2,0)$ scalar multiplet, on the other hand,
consists of $2n$ real scalar fields $\f^\a (\a=1,...,2n)$ and $2n$
Majorana fermions $\l^r\ (r=1,...,n\ {\rm and\ the }\ SO(2)\ {\rm
spinor\ indices\ are\ suppressed})$.

In \cite{mc}, the sigma model manifold $M$ was taken to be a coset
space of the form $G/H\times SO(2)$ where $G$ can be compact or
non-compact and $H\times SO(2)$ is the maximal compact subgroup of
$G$, where $SO(2)$ is the $R$-symmetry group. In particular, the
following cases are considered \cite{mc}

\be
M_+= {SO(n+2)\over SO(n)\times SO(2)}\ ,
\quad\quad
M_-= {SO(n,2)\over SO(n)\times SO(2)}\ .
\la{pm}
\ee

The results can be readily translated to the case of $G/H\times
U(1)$ with $G=SU(n+1)$ or $SU(n,1)$ and $H=SU(n)$. \\

Key ingredients in the description of the model are the matrices
$(L_I{}^i, L_I{}^r)$ where $I=1,...,n+2,\ i=1,2,\ r=1,..,n$, which
form a representative of the coset $M_\pm$. It follows that

\bea
&& L_I{}^i L^{Ij} = \pm \d^{ij}\ , \quad\quad L_I{}^r L^{Is}=\d^{rs}\ ,
\quad\quad L_I{}^i L^{Ir} = 0\ ,
\la{cons}\w2
&& \pm L_I{}^iL^{Ji} + L_I{}^r L^{Jr} = \d_I^J\ ,
\nn
\eea

where $\pm$ correspond to the scalar manifolds $M_{\pm}$. The $SO(n)$,
$SO(2)$ and $SO(n+2)$ vector indices are raised and lowered with the
Kronecker deltas and the $SO(n,2)$ vector indices with the metric
$\eta_{IJ}= {\rm diag}(++...+--)$.

Other important ingredient of the model is the $SO(2)$ gauged
pull-back of the Maurer-Cartan form on $M_\pm$ which is decomposed
into the $SO(n)\times SO(2)$ connections $Q_\m^{rs}$ and
$Q_\m^{ij}$, and the nonlinear covariant derivative $P_\m^{ir}$ as
follows:

\be
P_\m^{ir} = \left(L^{-1} D_\m L\right)^{ir}\ ,
\ \ \ \
Q_\m^{ij} = \left(L^{-1} D_\m L\right)^{ij}\ ,
\ \ \ \
Q_\m^{rs} =  \left(L^{-1} D_\m L\right)^{rs}\ ,
\la{mc}
\ee

where the $SO(2)$ covariant derivative is defined as

\be
D_\m L= \left( \del_\m +\ft12 A_\m^{ij} T_{ij}\right)L \ ,
\ee

The anti-hermitian $SO(2)$ generator $T_{ij}$ occurring in this
definition is realized in terms of an {\small $(n+2)\times (n+2)$}
matrix, which can be chosen as $(T_{ij})_I{}^J= (\,\pm\,
\d_{Ii}\,\d_{j}^J - i \leftrightarrow j\,) $. Introducing the
coordinates $\phi^\a\,(\a=1,...,2n)$ which parametrize the scalar
manifold $G/H$, we can also define the coset vielbein $V_\a^{ir}$
and the $SO(2)\times SO(n)$ connections $A_\a^{ij}, A_\a^{rs}$ on
$G/H$ as

\be
V_\a^{ir} = \left(L^{-1} \del_\a L\right)^{ir}\ ,
\ \ \ \
A_\a^{ij} = \left(L^{-1} \del_\a L\right)^{ij}\ ,
\ \ \ \
A_\a^{rs} =  \left(L^{-1} \del_\a L\right)^{rs}\ ,
\la{mc2}
\ee

where $\del_\a \equiv {\del\over \del\phi^\a}$. From the above
relations it follows that

\bea
P_\m^{ir} &=& \del_\m\phi^\a V_\a^{ir} + A_\m S^{ir}\ ,
\la{p1}\w2
Q_\m  &=& \del_\m\phi^\a A_\a+ A_\m C \ ,
\la{q1}\w2
Q_\m^{rs} &=& \del_\m\phi^\a A_\a^{rs} + A_\m C^{rs}\ ,
\la{q2}
\eea

where $A_\a^{ij}=A_\a \e^{ij}$, $A_\m^{ij}= A_\m \e^{ij}$,
$Q_\m^{ij} = Q_\m \e^{ij}$ and the $(C,S^{ir})$ functions are
defined as

\bea
\e_{ij}\e^{kl} C &=& (L^{-1}T_{ij}L)^{kl}\ ,
\nn\w2\
\e_{ij}C^{rs} &=& (L^{-1}T_{ij}L)^{rs}\ ,
\nn\w2 \e_{ij}S^{kr} &=& (L^{-1}T_{ij}L)^{kr}\ .
\la{cf}
\eea

The matter coupled $N=(2,0)$ Chern-Simons supergravity Lagrangian
which makes use of these ingredients has been obtained in \cite{mc}.
Up to quartic fermions the Lagrangian is as follows \cite{mc}:
%
\footnote{\footnotesize {Conventions: $\eta_{ab}=(-++), \
\eb=\ve^\dagger i\c_0$, $\c^\m C$ and $\c^{\m\n}C$ are symmetric and
$\c^{\m\n\r}={1\over \sqrt{-g}}\e^{\m\n\r}$. The $SO(2)$ charge
conjugation matrix is unity, \,$\C^i$\, is symmetric and
$\{\C^i,\C^j\}=2\d^{ij}$. A convenient representation is $\C_1=\s_1$,
$\C_2=\s_3$. We define $\C_3=\C_1\C_2$. Note that $(\C^3)^2=-1$.}}

\bea
e^{-1}{\cal L} &=&  {1\over 4} R + {e^{-1}\over 2} \e^{\m\n\r}\pb_\m D_\n \psi_\r
-{e^{-1}\over 16 ma^4}\, \e^{\m\n\r} A_\m \del_\n A_\r
-{1\over 4a^2} P_\m^{ir} P^\m_{ir}
\nn\w2
&& +{1\over 2}\,\lb_r\c^\m D_\m \l^r
   + {1\over 2a}\, \lb_r \c^\m\c^\n \C_i\psi_\m P_\n^{ir}
   -{m\over 2}\, \pb_\m\c^{\m\n}\psi_\n C^2
\nn\w2
&&  -2 ma\,\pb_\m \c^\m\C_i\C^3\l_r C S^{ir}
    -{1\over 2} m (1+ 4\e a^2)\, \lb^r\l_r C^2
\nn\w2
&& + {2 ma^2}\,\lb_r\C^3\l_s C^{rs} C
+ {2 ma^2}\,\lb_r\C_i\C_j\l_s S^{ir}S^{js}
\nn\w2
&& +{2 m^2}C^2(C^2 -2 a^2 S^{ir}S_{ir})\ ,
\la{a}
\eea

which has the local $N=2$ supersymmetry

\bea
\d e_\m{}^a &=& -\eb\c^a\psi_\m\ ,
\nn\w2
\d \psi_\m &=& D_\m \ve +m \c_\m C^2 \ve\ ,
\nn\w2
\d A_\m &=& 4 m a^2 \, (\eb \C^3\psi_\m)\, C
- 4 ma^3\,(\lb_r\c_\m\C_i\ve)\, S^{ir}\ ,
\nn\w2
L_i{}^I \d L_I{}^r &=&a  \eb\, \C_i \l^r\ ,
\nn\w2
\d\l^r &=& \left( -{1\over 2a} \c^\m P_\m^{ir}
+ 2 ma \C^3 C S^{ir}\right) \C_i\ve\ .
\la{s2}
\eea

The functions $C$ and $S^{ir}$ are defined in \eq{cf} and
\footnote{The $\e$ term in $D_\m \l^r$ was inadvertently omitted in
\cite{mc}.}

\bea
D_\m \ve &=& \left(\del_\m +\ft14 \o_\m{}^{ab} \c_{ab}
-{1\over 2 a^2 }\,Q_\m \C^3\right)\ve\ ,
\nn\w2
D_\m \l^r &=&\left(\del_\m +\ft14 \o_\m{}^{ab} \c_{ab}
+\left(\e+{1\over 2 a^2}\right)  Q_\m \C^3 \right)\l^r +Q_\m{}^{rs}\l^s\ .
\la{d}
\eea

The parameter $\e=\pm$1 corresponds to the manifolds $M_\pm$
defined in \eq{pm}, and the constant $a$ is the characteristic
curvature of $M_\pm$ (e.g. $2a$ is the inverse radius in the case
of $M_+=S^2$). The gravitational coupling constant $\k$ has been
set equal to one, but it can easily be introduced by dimensional
analysis. The constant $m$ is the $AdS_3$ cosmological constant.
Unlike in a typical anti de Sitter supergravity coupled to matter,
here the constants $\k,a,m$ are not related to each other for
non-compact scalar manifolds, while $a$ is quantized in terms of
$\k$ in the compact case as \cite{mc}.

To conclude this section, and for later purposes, we list the
equations of motion which follow from the Lagrangian \eq{a}:

\bea
&& R_{\m\n}- a^{-2} P_\m^{ir} P_\n^{ir} +8m^2 C^2\left(C^2
-2a^2 S^{ir}S_{ir}\right)\,g_{\m\n} \se 0\ ,
\la{e1}
\w3
&&\psi_{\m\n} +2m \c_{[\m}\,\psi_{\n]}\,C^2
+2ma\c_{\m\n} \C_i\C^3\l_r C S^{ir} -{1\over 2a} \C_i\c^\r\c_{\m\n}
\l_r P_\r^{ir}  \se  0\ ,
\la{e2}
\w3
&&F_{\m\n}-4ma^2 \sqrt{-g} \e_{\m\n\r} P^\r_{ir}S^{ir}\se 0 \ ,
\la{e3}
\w3
&&\c^\m D_\m \l^r  -m (1+ 4\e a^2)\, \l^r C^2
+{1\over 2a} \, \c^\m\c^\n \C_i\psi_\m P_\n^{ir}
+ {4ma^2} \C^3\l_s C^{rs} C
\nn
\w2
&&\qqq + {4ma^2} \C_i\C_j\l_s S^{ir}S^{js}
-2ma\c^\m\C^3\C_i \psi_\m C S^{ir}  \se 0\ ,
\la{e4}
\w3
&& D^\m P_\m^{ir} + 16 m^2 a^2\e_{ij}
S^{jr}C \left((1+ \e a^2)C^2-a^2\, S^{ks}S_{ks}\right)
\nn
\w2 &&
\quad\quad\quad +16 m^2 a^4\,C^{rs}S_{is}C^2\se 0\ ,
\la{e5}
\eea

where the fermion bilinears in the bosonic field equations have been
suppressed and

\bea
\psi_{\m\n} &=& D_\m \psi_\n - D_\n \psi_\m\ ,
\la{gfs}\w2
D_\m P^{\m ir} &=& {1\over \sqrt {-g}} \del_\m \left(\sqrt{-g}
g^{\m\n} P_\n^{ir}\right)+\e\, Q_\m^{ik}P^{\m\,kr} + Q_\m^{rs} P^{\m\,ir}\ .
\eea


\section{Connection with the Izquierdo-Townsend Model \label{secfour}}


The model reviewed above \cite{mc} differs from the one
constructed by Izquierdo and Townsend \cite{it}, in all the terms
containing the $C$ and $S$-functions. These differences stem from
the fact that the scalar fields in the model above are charged
under the R-symmetry group $SO(2)$ while in the model of \cite{it}
they are neutral. Given that this charge is related to the sigma
model radius, taking the zero charge limit in order to compare the
two models is expected to constrain the scalar manifold. Here, we
will show the relation between the two models and show that they
indeed agree only in the limit in which the scalar  manifold is
flat.

We begin by parametrizing the coset representative $L$ as
follows

\be
L=exp \left(\matrix{0 & \f^{ir} \cr -\e (\f^{ir})^T & 0\cr}\right)\ ,
\la{cr}
\ee

where $\phi^{ir}$ are $2n$ real coordinates on $M_\pm$. Next, we
perform the rescalings

\be
A_\mu \rightarrow a^2\, A_\mu\ , \quad\quad \phi^{ir} \ra  a\,
\phi^{ir}
\ee

and consider the limit $a^2 \rightarrow 0$. From the definitions
in \eq{mc} we find

\bea
P_\mu^{ir} & = & a \partial_\mu \phi^{ir} + \cdots\ ,
\nn\w2
Q_\mu & = & a^2 \left( A_\mu - {1 \over 2} \phi^{ir} \partial_\mu
\phi_r{}^j \epsilon_{ij} \right) + \cdots\ ,
\nn\w2
Q^{rs}_\mu & = & a^2 ( \partial_\mu \phi^{ri} \phi_i{}^s
- \phi^{ri} \partial_\mu \phi_i{}^s ) + \cdots\ ,
\eea

where $\cdots$ denote higher order terms in positive powers of $a^2$. Let us define

\be
d\phi^\alpha A_\alpha
= {1 \over 2} \phi^{ir} d \phi^j{}_r\, \epsilon_{ij}\ ,
\ee

so that $Q_\mu = a^2(A_\mu + \partial_\mu \phi^\alpha A_\alpha)$,
where the index $\alpha$ represents a pair of indices $(ir)$. We
have

\be
{1 \over 2} d\phi^\alpha \wedge d\phi^\beta F_{\alpha\beta}
= {1 \over 2} d\phi^{ir} \wedge d\phi^j{}_r\, \epsilon_{ij}\ ,
\la{fab}
\ee

where $F_{\alpha\beta} = \partial_\alpha A_\beta
- \partial_\beta A_\alpha$. In the limit $a^2 \rightarrow 0$ the Lagrangian becomes

\bea
e^{-1}{\cal L} &=&  {1\over 4} R
 + {1\over 2} e^{-1} \e^{\m\n\r}\pb_\m D_\n \psi_\r
-{1\over 16}{e^{-1}\over m }\, \e^{\m\n\r} A_\m \del_\n A_\r
-{1\over 4} \del_\m \phi^{ir} \del^\m \phi_{ir}
\nn\w2
&& +{1\over 2}\,\lb_r\c^\m D_\m \l^r
   + {1\over 2}\, \lb_r \c^\m\c^\n \C_i\psi_\m \del_\nu \phi^{ir}
\nn\w2
&&
-{1 \over 2} m \, \pb_\m\c^{\m\n}\psi_\n
-{1\over 2} m \,\lb^r\l_r +{2 m^2}\ ,
\la{a3}
\eea

and the transformation rules become

\bea
\d e_\m{}^a &=& -\eb\c^a\psi_\m\ ,
\nn\w2
\d \psi_\m &=& D_\m \ve +m \c_\m \ve\ ,
\nn\w2
\d A_\m &=& 4 m \, \eb\, \C^3\psi_\m \ ,
\nn\w2
\d \phi^{ir} &=& \eb\, \C^i \l^r\ ,
\nn\w2
\d\l^r &=& -{1\over 2} \c^\m \del_\mu \phi^{ir} \C_i\ve \ ,
\la{s3}
\eea

where the covariant derivatives are defined by

\bea
D_\m \psi_\nu &=& \left(\nabla_\m -{1\over 2}\,\del_\m\phi^\a A_\a \C^3
-{1\over 2}\,A_\m \C^3\right) \psi_\nu\ ,
\nn\w2
D_\m \l^r &=&\left(\nabla_\m +{1\over 2} \del_\m\phi^\a A_\a  \C^3
+{1\over 2} A_\m \C^3\right)\l^r\ ,
\nn\w2
D_\m \ve &=& \left(\nabla_\m -{1\over 2}\,\del_\m\phi^\a A_\a \C^3
-{1\over 2}\,A_\m \C^3\right)\ve\ ,
\la{d3}
\eea

and $\nabla_\m =\del_\m +\ft14 \o_\m{}^{ab} \c_{ab}$. Introducing

\bea
A'_\mu & = & A_\mu + \del_\m\phi^\a A_\a
\nn\w2
& = & A_\mu + {1 \over 2} \phi^{ir} \partial_\mu \phi^j{}_r
\epsilon_{ij}\ ,
\la{defa}
\eea

the Lagrangian \eq{a3} becomes

\bea
e^{-1}{\cal L} &=&  {1\over 4} R
+ {1\over 2} e^{-1} \e^{\m\n\r}\pb_\m D_\n \psi_\r
-{1\over 16}{e^{-1}\over m }\, \e^{\m\n\r} A'_\m \del_\n A'_\r
-{1\over 4} \del_\m \phi^{ir} \del^\m \phi_{ir}
\nn\w2
&& -A'_\m J^\m + {1\over 2} \del_\m \phi^\a A_\a J^\mu
   +{1\over 2}\,\lb_r\c^\m D_\m \l^r
   + {1\over 2}\, \lb_r \c^\m\c^\n \C_i\psi_\m \del_\nu \phi^{ir}
\nn\w2
&& -{1 \over 2} m \, \pb_\m\c^{\m\n}\psi_\n
-{1\over 2} m \,\lb^r\l_r + {2 m^2}\ ,
\la{a4}
\eea

where the current $J^\m$ is defined as

\be
J^\m =  -{1 \over 16me} \epsilon^{\mu\nu\rho}
\partial_\nu \phi^\alpha \partial_\rho \phi^\beta F_{\alpha\beta}\ ,
\ee

with $F_{\a\b}$ defined in \eq{fab}. The transformations rules
\eq{s3}, on the other hand, become

\bea
\d e_\m{}^a &=& -\eb\c^a\psi_\m\ ,
\nn\w2
\d \psi_\m &=& D_\m \ve +m \c_\m \ve\ ,
\nn\w2
\d A'_\m &=& 4 m \, \eb\, \C^3\psi_\m
- \delta\phi^\alpha \partial_\mu \phi^\beta F_{\beta\alpha}
\nn\w2
\d \phi^{ir} &=& \eb\, \C^i \l^r\ ,
\nn\w2
\d\l^r &=& -{1\over 2} \c^\m \del_\mu \phi^{ir} \C_i\ve \ ,
\la{s4}
\eea

where we have discarded a term in $\d A_\m$ which can be expressed
as a gauge transformation $\d A_\m = \del_\m \L$. Of course, the
above transformations are up to cubic fermion terms in the
transformation rules of $\psi_\m$ and $\l^r$, since the Lagrangian
\eq{a4} is up to quartic fermion terms. Note also that the covariant
derivatives have now simplified to

\bea
D_\m \psi_\nu &=& \left(\nabla_\m -{1\over 2}\,A'_\m \C^3\right)\psi_\nu\ ,
\nn\w2
D_\m \l^r &=&\left(\nabla_\m +{1\over 2} A'_\m \C^3\right)\l^r\ ,
\nn\w2
D_\m \ve &=& \left(\nabla_\m -{1\over 2}\,A'_\m \C^3\right)\ve\ .
\la{d4}
\eea

The formulae \eq{a4}, \eq{s4} and \eq{d4} agree with those of the
Izquierdo-Townsend model \cite{it} for the flat sigma model. Note
that in trying to set the U(1) charge of the scalar fields equal
to zero, we have been forced to flatten the sigma model manifold.
This is due to the fact that the U(1) charge is related to the
radius of the scalar manifold. The flat model discussed here will
be used in Section 5.3.


\section{Boundary Conditions and Linearized Field Equations \label{secfive} }


In order to examine the properties of the model described above
near the boundary, we shall begin by fixing certain gauges and
studying the behaviour of the linearized field equations near the
boundary.

The $AdS_3$ spacetime can be covered by two regions each of which is
parametrized by a set of Poincar\'e coordinates $(x^0,x^1,x^2)$ in
${\bf R}^3$ with $x^2>0$. We shall use the notation $x^\mu =(x^0,x^1)$ and
$x^2 \equiv r$. This patch contains half the boundary of $AdS_3$ in the
form of the Minkowskian plane at $r=0$. The other region is behind the
horizon at $r=\infty$. In what follows we shall work only within one of
the regions.

Following \cite{t1}, we choose the following gauge conditions

\bea
&& e_r{}^2= {1\over 2mr}\ , \quad e_r{}^a=0\ , \quad e_\m{}^2=0\ ,
\nn\w2
&&\psi_r=0, \quad\quad A_r =0\ ,
\la{gc}
\eea

where $a=0,1$ is the tangent space index in $D=2$. Note that the
second coordinate is labeled as $r$ in curved space and as $2$ in
tangent space. The metric in this gauge takes the form

\be
ds^2 = {1\over (2mr)^2} (dr^2 + dx^\m dx^\n \hat g_{\m\n} )\ ,
\la{m}
\ee

where $\hat g_{\m\n}= \hat e_\m{}^a \hat e_\n{}^b \eta_{ab}$. The
SO(2,2) invariant AdS metric corresponds to the case $\hat
g_{\m\n}= \eta_{\m \n}$. The components of the spin connection
following from the metric \eq{m} are

\bea
\omega_\mu{}^{ab} & = & \hat \omega_\mu{}^{ab}\ ,
\qquad
\omega_r{}^{ab} = - \hat e^{[b\nu} \partial_r \hat e_\nu{}^{a]}\ ,
\nn\w2
\omega_\mu{}^{a2} & = & - {1 \over r} \hat e_\mu{}^a + \hat e^{(b\nu}
\partial_r \hat e_\nu{}^{a)} \hat e_{\mu b}\ , \qquad \omega_r{}^{a2} = 0\ .
 \eea

When $\hat g_{\mu\nu} = \eta_{\mu\nu}$, the only nonvanishing component is
$\omega_\mu{}^{a2} = - \delta_\mu^a / r$.

We next study the asymptotic behaviour of the solutions of the
linearized field equations near the boundary $r=0$. We are going to
do this in Euclidean signature. In this signature the AdS space
consists of a single region covered by Poincar\'e coordinates plus a
point at $r=\infty$. This point is actually a boundary point and the
boundary has the topology of the two sphere, represented in the
Poincar\'e coordinates by the Euclidean plane at $r=0$ plus the
point at infinity.

We will assume that the dreibein $e_\m{}^a$ behaves as $r^{-1}$ as
in the $SO(2,2)$ invariant case. To determine the asymptotic
behaviours of the remaining fields, we need to examine their linearised field
equations as expanded around the supersymmetric $AdS$ background in
which the only nonvanishing fields are
\be
\hat g_{\m\n}=\eta_{\m\n}\ ,\quad\quad L=1\ .
\la{bg}
\ee

Next, we use the coset representative $L$ given in \eq{cr} which
leads to the following expressions at the linearized level

\be
P_\m ^{ir} \ra \del_\m \f^{ir}, \quad C \ra 1,
\quad C^{rs} \ra 0, \quad S^{ir} \ra \e \ \e ^{ij} \f ^{jr}\ .
\ee

The field equations for $A_\m$ and $\psi_\m$ linearized around the
background \eq{bg} are
\footnote{We use the convention $\c^{\m\n}=
{1\over \sqrt{-g}}\e^{\m\n}\c^2$.}
\be
\del_r A_\m \se 0\ , \qq \del_{[\m}A_{\n]}\se 0\ ,
\la{ve}
\ee
\be
\del_r \psi_{\m\pm} \se \mp \ft12 r^{-1}\,
\psi_{\m\pm}\ .
\la{sg}
\ee

where the suffixes $\,\pm\,$ indicate the eigenvalues of $\c^2$, which
in turn indicate the chiralities of the spinors on the boundary. The
equations involving radial derivatives are readily solved to all orders
in $r$, and in a convenient normalization we have

\be
A_\mu \se A_{(0)\m} \ , \qq \psi_\m \se (2mr)^{-\ft12}\,\psi_{\m(0)+}
+(2mr)^{\ft12}\,\psi_{\m(0)-} \ .
\la{1}
\ee

To give the proper boundary condition for the vector
field, we define
\footnote{We use a notation in which the chiralities and Hodge
dualities are labeled by lower $\pm$ indices and the regular and
irregular nature of boundary conditions are labeled by upper
$\pm$ indices.}

\be
A_{(0)\m\pm} \se P_{\m\pm}{}^\n A_{(0)\n}
\se {1\over 2} (g_{\m\n} \pm \sqrt{-g} \e_{\m\n})A_{(0)}^\n
\ .\la{hodge}
\ee

The remaining equation for $A_{\m}$ in \eq{ve} then amounts to

\be
\del^{\m}_-A_{\m+} \se \del^{\m}_+A_{\m-}\ .
\la{ve2}
\ee

As we shall show in the next section, the anti self dual component
$A_{(0)\m-}$ forms an off-shell $d=2$ supermultiplet together with
$\psi_{(0)\m+}$ and $e_{(0)\m}{}^{a}$. Thus it is natural to treat
$A_{(0)\m-}$ as the independent boundary field, and let $A_{(0)\m+}$ be
determined from \eq{ve2}. The fact that only one of the Hodge dualities of the
vector field is independent can also be understood by considering the
Hamiltonian formulation of the bulk Chern-Simons theory, where
$A_{(0)\m\pm}$ form a pair of canonically conjugate variables. Thus, the
proper boundary conditions for the supergravity multiplet are:

\be
e_\m{}^a \sim (2mr)^{-1} e_{(0)\m}{}^a\ , \qq \psi_{\m+} \sim
(2mr)^{-\ft12}\,\psi_{(0)\m +} \qq A_{\mu-} \sim A_{(0)\m-}\ .\la{b1}
\ee

We now turn to the discussion of the boundary conditions on the
matter fields, starting with the scalar fields.

\subsection{Matter Scalars}

The linearized scalar field equation near the boundary is given
by (the $r$-dependence is shown explicitly and the $SO(2)\times
SO(n)$ indices of $\phi^{ir}$ and $\l^r$ are suppressed):

\be
r^2\del^\m \del_\m \phi+ r^3\del_r\left( r^{-1} \del_r \phi \right)
- m_{\f}^2\phi = 0\ ,
\la{se}
\ee

where

\be
m_{\f}^2 = 4a^2(a^2 + \e)= \Delta(\Delta-2)\ ,
\ee

and $\Delta$ equals $\Delta_+$ or $\Delta_-$ defined by

\be \D_{\pm}(\f) = 1 \pm \sqrt {1+ m_{\f}^2}\ .
\la{cw}
\ee

Thus, in terms of $\e$ and $a^2$:

\bea
\e\se 1&:& \mx{\{}{l}{\D_+(\f) \se 2+2a^2\ ,\\ \\
\D_-(\f) \se -2a^2\ ,}{.}\w3
\e\se -1 &:& \mx{\{}{l}{\D_+(\f)\se 1+|1-2a^2|\ ,\\ \\
\D_-(\f)\se 1-|1-2a^2|\ .}{.}
\la{ds}
\eea

A free scalar field $\phi$  behaves near the boundary as

\be
\f(r,\vx)\sim (2mr)^{\D_-(\f)}\left[\phi_{(0)}^+ +
(2mr)^2\phi_{(2)}^+ +\cdots\right] +(2mr)^{\D_+(\f)}\left[\phi_{(0)}^-
+ (2m r)^2 \phi_{(2)}^- + \cdots\right]
\la{bc}
\ee

for $2 a^2 \notin {\bf Z}$, and

\be
 \f(r,\vx) \sim (2mr)^{\D_-(\f)}\left[\f_{(0)}^+ +
(2mr)^2\f_{(2)}^+ +\cdots \right]
+(2mr)^{\D_+(\f)} {\rm ln}\, (2mr)\ \left[\f_{(0)}^-
+ (2mr)^2 \f_{(2)}^- + \cdots\right]
\la{bc2}
\ee

for $ 2 a^2 \in {\bf Z}$. The expansion coefficients
$\f_{(2n)}^\pm$ depend only on $\vx$. For $2a^2\notin{\bf Z}$ and
for $\e=-1,a^2=\ft12$, that is $\D_+=1$, the coefficients
$\f^\pm_{(2n)}$, $n\ge 1$, are determined through the linearized
field equations as local expressions in terms of $\f_{(0)}^\pm$.
For other values of $\e$ and $a^2$, that is for $\D_+=2,3,...$,
the coefficients $\f^+_{(2n)}, n\leq \D_+-2$ are given in terms of
$\f^+_{(0)}$ while $\f^+_{(2\D_+-2)}$ is undetermined and thus
independent. At higher order in $r$ one then finds that
$\f^-_{(0)},\f^-_{(2)},\dots$ are given in terms of $\f^+_{(0)}$
and that $\f^+_{(2\D_+)},\f^+_{(2\D_++2)},\dots$ are given in
terms of $\f^+_{(2\D_+-2)}$. The above results follow from the
small $z$ expansion of the modified Bessel functions \cite{mr}.
Similar results hold for small perturbations around the anti-de
Sitter background \cite{2}.

There are two types of boundary conditions that may be imposed on
the scalars: regular conditions which amount to specifying the
leading component at the boundary and irregular conditions which
amount to specifying the independent subleading component
described above \cite{bkl,kw}, and which are possible when $\D_-
\ge 0$. Thus, for $2a^2\notin {\bf Z}$, a regular boundary
condition amounts to specifying $\f_{(0)}^+$ while an irregular
boundary condition amounts to specifying $\f_{(0)}^-$. For
$\D_+=1$, that is, for $\e=-1,a^2=\ft12$, a regular boundary
condition amounts to specifying $\f^-_{(0)}$ and an irregular
condition amounts to specifying $\f^+_{(0)}$.

Given that $\phi_{(0)}^{\pm}$ are associated with conformal
operators of weight $\Delta_{\pm}(\f)$, the requirement of
unitarity imposes the following restriction for irregular
boundary conditions:

\be
\Delta_{-}(\f) \ge 0\ .
\la{uc}
\ee

For regular boundary conditions, the unitarity condition is
automatically satisfied, while for irregular conditions \eq{uc}
restricts the possible values of $a^2$. Thus it follows that the
following boundary conditions are possible:

\bea
\e= 1&:& \mbox{regular:}\quad\ \ \f\sim(2mr)^{\D_-(\f)}\f^+_{(0)}
\quad \mbox{for $a^2\ge 0$}
\w4
&&\mbox{irregular:}\quad \f\sim (2mr)^2\f^{+}_{(2)}
\quad \mbox{for $a^2=0$}\ .
\w4
\e= -1 &:& \mbox{regular:}\quad\
\f\sim\mx{\{}{ll}{(2mr)^{\D_-(\f)}\f^+_{(0)}&\mbox{for $a^2\ne \ft12$}\ ,\\ \\
(2mr){\rm ln}(2mr)\f^{-}_{(0)}&\mbox{for $a^2=\ft12$}\ .}{.}
\la{sbc1}
\w4
&&\mbox{irregular:}\quad \f\sim\mx{\{}{ll}{(2mr)^{\D_+(\f)}\f^-_{(0)}&
\mbox{for $0< a^2<1$}\ ,\ a^2\ne \ft12\ ,\\ \\
(2mr) \f_{(0)}^{+}& \mbox{for $a^2=\ft12$}\ ,\\ \\
(2mr)^2\f^{+}_{(2)}&\mbox{for $a^2=0,1$}\ .}{.} \la{sbc2} \eea

\subsection{Matter Fermions}

We now turn to the boundary conditions on the matter fermions. The
linearized equations obeyed by the matter fermions near the boundary is

\be
r\c^\m \del_\m \l +r\c_2 \del_r\l -\c_2 \l -m_{\l}\l =0\ ,
\la{fe}
\ee

where the fermion mass is given by

\be
 m_{\l} = {1 \over 2} (1 + 4\e a^2)\ .
\ee

We find that for $m_\l\notin {\bf Z} +\ft12$ a solution to \eq{fe}
is given by

\be
\l\se (2mr)^{1-m_{\l}}\left[\l_{(0)-}+2mr\l_{(1)+}+\cdots\right]
+(2mr)^{1+m_{\l}}\left[\l_{(0)+}+2mr\l_{(1)-}+\cdots\right]\ ,
\la{fex1}
\ee

where $\pm$ refers to the $\c^2$ eigenvalue. For $m_\l=\ft12,
\ft32,\ft52,...$, the solution takes the form

\be
\l = (2mr)^{1-m_{\l}}\left[\l_{(0)-}+2mr\l_{(1)+}+\cdots \right]
+(2mr)^{1+m_{\l}}\,{\rm ln}\,2mr\ \left[\l_{(0)+}
+2mr\l_{(1)-}+\cdots\right]\ .
\la{fex2}
\ee

For $m_\l=-\ft12,-\ft32,-\ft52,...$, the solution is given by
\eq{fex2} with $m_\l \ra -m_\l$ and all the chiralities flipped. For
later use, we also record the following relation

\be
\l_{(1)\pm} \se {1\over 2m(2m_\l\mp 1)} \c^\m \del_\m
\l_{(0)\mp}\ , \quad 2m_\l\mp1 \neq 0\ .
\la{sl}
\ee

Note that unlike for the scalars, the coefficients in the
logarithmic branch in \eq{fex2} is never undetermined. Next,
following \cite{hs}, we define the conformal weights of the
fermions as

\be
\D_{\pm}(\l) \se 1\pm |m_\l| \se  1\pm \ft12|1+4\e a^2|\ .
\la{fw}
\ee

Thus, in terms of $\e$ and $a^2$:

\bea
\e\se 1&:& \mx{\{}{l}{\D_+(\l) \se \ft32+2a^2\,\\ \\
\D_-(\l) \se \ft12-2a^2\ .}{.}\w3
\e\se -1 &:& \mx{\{}{l}{\D_+(\l)\se 1+\ft12|1-4a^2|\ ,\\ \\
\D_-(\l)\se 1-\ft12|1-4a^2|\ .}{.}\la{df}
\eea

Thus, the regular boundary conditions are associated with
$r^{\D_-(\l)}$ behaviour and the irregular boundary conditions
with $r^{\D_+(\l)}$ behaviour (note that this holds for all
values of $a^2$). It follows from \eq{fex1} and \eq{fex2} that the
chirality of the regular and irregular boundary spinors
$\l^{\pm}_{(0)}$ is equal to minus the sign of the fermion mass:

\be
\left(\c^2\pm{m_\l\over |m_\l|}\right) \l_{(0)}^{\pm} \se 0 \ ,
\la{fc}
\ee

where the superscript $``+"$ refers to regular and $``-"$ to
irregular boundary conditions. Thus, in the case of regular
boundary conditions the chirality is negative for positive
fermion mass, and positive for negative fermion mass. In the case
of irregular boundary conditions the chirality is positive for
positive mass and negative for negative mass.

Imposing the following unitarity condition

\be \Delta_-(\l)\geq \ft12\ ,\la{dl} \ee

it follows that the allowed boundary conditions for the matter
fermions are:

\bea \e=1 &:& \qquad \mbox{regular:}\quad \l\sim
(2mr)^{\D_-(\l)}\l_{(0)-}\quad \mbox{for $a^2 \geq 0$}\ ,
\w3
&& \quad \mbox{irregular:}\quad \l\sim
(2mr)^{\ft32}\l_{(1)+}\quad \mbox{for $a^2=0$}\ .
\w5
\e=-1 &:& \quad \mbox{regular:}\quad \l\sim\mx{\{}{ll}{
(2mr)^{\D_-(\l)}\l_{(0)-}& \mbox{for $0\leq a^2\le \ft14$}\ ,\\ \\
(2mr)^{\D_-(\l)}\l_{(0)+}& \mbox{for $a^2\ge \ft14$}\ .}{.}
\la{fbc1} \w3 &&\quad \mbox{irregular:}\quad \l\sim\mx{\{}{ll}{
(2mr)^{\ft32}\l_{(1)+}&\mbox{for $a^2=0$}\ ,\\ \\
(2mr)^{\D_+(\l)}\l_{(0)+}&\mbox{for $0<a^2<\ft14$}\ , \\ \\
(2mr)^{\D_+(\l)}\l_{(0)-}&\mbox{for $\ft14 < a^2<\ft12$}\ ,\\ \\
(2mr)^{\ft32}\l_{(1)-}&\mbox{for $a^2=\ft12$}\ .}{.} \la{fbc2}
\eea

Note that for $\e=-1$ and $a^2=\ft14$ the regular boundary
condition can be imposed on either of the chiralities.


\section{The Local Conformal Supersymmetry on the Boundary
\label{secsix} }


In this section we shall derive the realization of the $d=2$,
$N=(2,0)$ conformal supersymmetry on the boundary supergravity
multiplet $(e_{(0)\m}{}^a, \psi_{(0)\m +}, A_{(0)\m-})$ and
boundary chiral multiplets involving fields that are to be
specified case by case in accordance with the boundary conditions.

The $d=2$ symmetries are found by examining the nature of the bulk
transformation rules close to the boundary. To analyze this we
first find the $D=3$ transformation parameters which preserve the
$D=3$ gauge conditions \eq{gc} near the boundary. We then evaluate
the resulting $D=3$ transformations of a solution to the $D=3$
field equations with given set of boundary data $\Phi_{(0)}$. By
matching powers of $r$ in the limit when $r\ra 0$ we thus obtain
the resulting $d=2$ transformations $\d\Phi_{(0)}$. In specifying
the boundary data we have to choose between regular and irregular
boundary conditions such that $\delta \Phi_{(0)}$ is a local
expression in terms of $\Phi_{(0)}$ and its derivatives.


For $\e=1$ and $a^2>0$, the scalar fields diverge at the boundary
and the perturbative expansion breaks down, therefore we shall
exclude the case $\e=1$ from now on. For $\e=-1$ and $a^2>0$, the
matter scalars diverge for $a^2>1$ and the matter fermions diverge
for $a^2>\ft34$. Moreover, for $\ft12<a^2<\ft34$, the
supersymmetry transformation of the vector field involve matter
contributions which diverge as $r\ra 0$ (see footnote below
\eq{ld}), which is not consistent with its $r$ expansion. For
$0<a^2\le \ft12$ the $r$ expansion is well-defined, and in fact,
since all matter fields have positive Weyl weight for this case,
the nonlinearities vanish at the boundary. Finally, for $a^2=0$,
the appropriate model to consider is the ${\bf R} ^{2n}$ sigma
model, in which case the scalars have Weyl weight zero. In
summary, the perturbative expansion makes sense only for

\be \e=-1\ ,\quad 0 \le  a^2 \le \ft12\ , \qquad  \mbox{or} \quad
\e=+1\ , \quad a^2=0\ . \ee

We remark that in the case of single scalar multiplet coupling,
namely when the sigma model manifold is $S^2$ for $\e=1$ and $H^2$
for $\e=-1$, the excluded range of the parameter $a^2$ coincides
with the fact that the scalar potential has the form of a
confining well. In the allowed range, however, the potential is
unbounded from below. It would be interesting to study if the
potential exhibits qualitatively similar behaviour for arbitrary
number of scalar multiplets.


Importantly, regularity of the $D=3$ solutions determines the
irregular boundary conditions in terms of the regular, or vice
versa, which leads to the usual interpretations of the anti-de
Sitter/conformal field theory duality. This leads to subtleties,
however, in the case of irregular boundary conditions in the
matter sector, where the nonlinearities appear to lead to mixings
between the regular and irregular fields in the transformations of
the irregular fields (at least for certain rational values of
$a^2$). As a first step towards understanding this, it is
reasonable to begin by examining the nature of the transformations
of the irregular fields among themselves by formally putting the
regular fields to zero in the case of irregular boundary
conditions. This is our approach when $0<a^2<\ft12$, which we
refer to as case 1 below. For the special values $a^2=\ft12$ and
$a^2=0$, which we treat separately as case 2 and 3 below, the
nonlinearities are however more manageable, and for these two
cases we therefore keep both regular and irregular fields. Thus,
in summary, the boundary conditions in the matter sector are taken
as follows:

\be
\begin{array}{lcl} \mbox{Case 1:}& \e=-1\ ,\quad 0<a^2<\ft12&
\mbox{Either regular or irregular matter fields.}\\ \\
\mbox{Case 2:}& \e=-1\ ,\quad a^2=\ft12& \mbox{Both regular and
irregular
matter fields.}\\ \\
\mbox{Case 3:}& a^2=0& \mbox{Both regular and irregular matter
fields.}
\end{array}
\ee



\subsection{ Case 1\,:\ \ $\e=-1$ and $0<a^2<\ft12$ }


We begin by determining the asymptotic behaviors of the local symmetries
of the bulk which preserve the gauge conditions \eq{gc} near the boundary.
Using the asymptotic behavior of the supergravity multiplet fields given
in \eq{b1}, we find (without linearizing in the fields) that the residual
gauge symmetries are

\bea
\xi^r &=& -r \L_{D(0)}\ ,\quad
\xi^\m \se \xi_{(0)}^\m +\cO(r^2)\ ,
\nn\w2
\L_L^{a2} &=& \cO(r^2)\ ,\quad
\L_L^{ab} \se \L_{L(0)} \e^{ab} +\cO(r^2)\ ,
\nn\w2
\L &=& \L_{(0)} + \cO(r^{4a^2})\ ,
\nn\w2
\ve_\pm &=& (2mr)^{\mp\ft12} \left[ \ve_{(0)\pm} +\cO(r^2)\right]\ ,
\la{p}
\eea

where $\L_{D(0)}$, $\xi_{(0)}$, $\L_{L(0)}$, $\L_{(0)}$ and
$\ve_{(0)\pm})$ denote the parameters of dilatation,
reparametrization, Lorentz rotation, $SO(2)$ rotation and
supersymmetry, respectively, and the fields with suffix $(0)$ are
arbitrary functions of $x^\m$. Note that the parameter $\xi^r$ is
determined fully and it has only linear $r$-dependence, while the
other parameters have series expansions in $r$. The form of
$\x^r$, $\x^\m$ and $\L_L^{a2}$ come from the variations of the
gauge conditions involving the dreibein and the form of
$\L_L^{ab}$ can be deduced from the requirement of residual
Lorentz transformations on the boundary. The last two results
come from the variation of the gauge conditions $A_r=0$ and
$\psi_r=0$, respectively.

To derive the bosonic transformations in $d=2$, we insert \eq{p}
together with the expansions \eq{b1}, \eqs{bc}{bc2} and
\eqs{fex1}{fex2} into the bosonic transformations in $D=3$. These
do not mix different chiralities and powers of $r$. It is
therefore straightforward to read off the transformation for the
leading components. We find the usual general coordinate
transformations of all the fields with parameter $\xi^\m_{(0)}$,
and (in the rest of subsection we have dropped the $(0)$ labels
for notational simplicity)

\bea
\d e_{\m}{}^a &=& \left(\L_D\eta^{ab}-\L_L \e^{ab}\right)
e_{\m b}\ ,
\nn\w2
\d \psi_{\m +} &=& \left(\ft12 \L_D +\ft12 \L_L
+{1\over 2a^2} \L\C_3\right) \psi_{\m +}\ ,
\nn\w2
\d A_{\m-} &=&
\ft12 \left(g_{\m\n} -\sqrt{-g}\e_{\m\n}\right)\del^\n \L\ ,
\nn\w2
\d \phi^{\pm ir} &=& -\D_{\mp}(\f) \L_D \phi^{\pm ir}+ \L \e^{ij} \phi^{\pm jr}\ ,
\nn\w2
\d\l^{\pm r} &=& \left[-\D_{\mp}(\l)\L_D +\ft12\L_L\c_2
                    +(1-{1\over 2a^2}) \L\C_3 \right]\,\l^{\pm r}\ .
\la{bt}
 \eea

Note, the superscripts on the matter fields refer to
regular/irregular boundary conditions, and the chiralities of the
matter fermions are given by \eq{fc}.

To find the $d=2$ supersymmetry transformation rules we
substitute the expression for the supersymmetry parameter given in
\eq{p} into the $D=3$ supersymmetry transformation rules \eq{s2}
and take the limit $r\ra 0$. In the supergravity sector we find:

\bea
\d e_\m{}^a &=& -\eb \c^a \psi_\m \ ,
\nn\w2
\d \psi_\m &=& D_\m\ve + 2\c_\m \eta\ ,
\la{csg}
\w2
\d A_{\m-} &=& \ft12 a ^2\eb \C_3\c_\m\c^{\r\s}\psi_{\r\s} +
2a^2 {\bar\eta}\C^3\c^\n\c_\m \psi_\n \ ,
\nn
\eea

where we have introduced the notation

\bea
\c^\m&=&\c^a e_a{}^\m\ ,
\nn\w2
\ve &=& \ve_{+}\ ,
\nn\w2
\eta &=& m\ve_{-} -{1\over 8a^2}\,\C_3 \c^\m A_{\m+} \ve_{+}\ ,
\nn\w2
D_\m\ve &=& \left(\del_\m -\ft12\o_\m -{1\over 2a^2}\, A_{\m-}\C_3\right)\,\ve\
\la{nota}
\eea

and the $d=2$ gravitino field strength $\psi_{\m\n}$ is
defined as in \eq{gfs} but with the covariant derivative defined
above. We note the correction to the special supersymmetry
parameter $\eta$. In the gravitino transformation rule, this
correction arises from the $A_{\m+}$ contribution to the three
dimensional $D_\m\ve$, while in the vector transformation rule it
arises from the $D=3$ covariant derivative in the gravitino field
strength and from the varying the self-duality projector
according to the following:

\be
\d A_{\m\pm}\equiv \d(P_{\m\pm}{}^\n A_\n)\se P_{\m\pm}{}^\n\d A_\n \mp\bar{\ve}
\c_\m \psi^\n A_{\n+}\ ,
\la{da}
\ee

where the projection is defined in \eq{hodge}. We also note
that in obtaining the vector transformation rule we have
eliminated the anti self dual component of $\psi_{\m-}$ using
the boundary limit of the $\m\n$ component of the $D=3$ gravitino
equation \eq{e2} as follows:

\be
{m\over 2}\c_\m\c^\n\psi_{\n-} \se \ft14 \c^\n\psi_{\m\n}
+{1\over 8a^2}\C^3\c_\m\psi^\n A_{\n+}+ ma \c_\m\C^i\l^r_{+}\f^{+ir}\ ,
\la{fm}
\ee

where $\psi_{\m\n}=D_\m\psi_\n -D_\n\psi_\m$ with the covariant
derivative defined as in \eq{nota}. In deriving \eq{fm} one
notices that the last two terms in the $\mu\nu$ component of the
$D=3$ gravitino equation \eq{e2} add up in the leading order,
which follows from the fact that for $0<a^2<\ft12$ the $D=3$
scalar fields obey

\be
r\del_r \f^{ir}\se 2a^2(\f^{ir}+{\cal O}(r))\ .
\la{ld}
\ee

In the $r\m$ component of the $D=3$ gravitino equation, however,
these terms cancel, which means that there is no matter
contribution to the leading order. Thus we have obtained a local
realization of the boundary supersymmetry on the $(2,0)$
conformal supergravity multiplet in $d=2$ which is off-shell and
decoupled from matter
\footnote{At this point we can see why the range
$\ft12<a^2<\ft34$ does not yield a local realization of the
boundary supersymmetry: although the matter fields vanish at the
boundary, so that we can trust the $r$-expansion, \eq{ld} is now
replaced by $r\del_r\f^{ir} =(2-2a^2)(\f^{ir}+{\cal O}(r))$ which
results in divergent, matter dependent contributions to $\d
A_{\m-}$.}.

We next study the transformations of the matter fields. We recall
from \eqs{sbc1}{sbc2} and \eqs{fbc1}{fbc2} that both regular and
irregular boundary conditions are admissible for $0<a^2<\ft12$.
There are two sets of combined regular and irregular boundary
conditions which lead to two types of conformal $(2,0)$
supermultiplets which will be referred to as Type 1 and Type 2.
The Type 1 multiplet consists of regular scalars, and fermions
which are regular for $0<a^2<\ft14$ and irregular for
$\ft14<a^2<\ft12$. The Type 2 multiplet, on the other hand, is a
novel  multiplet which consists of irregular scalars, and
fermions which are irregular for $0<a^2<\ft14$ and regular for
$\ft14<a^2<\ft12$.

To find the transformations at the linearized level, we insert
\eq{b1}, \eq{bc} and \eq{fex1} into the $D=3$ transformation rules
in \eq{s2} and match powers of $r$. In the case of Type 2, we
also use the Dirac equation \eq{sl} in obtaining the scalar
transformation. The results are as follows:

\bea
\hspace{-2cm}\mbox{Linearized Type 1:}&&\begin{cases}{
\d\phi^{+\,ir} \se a \eb_+\,\C^i \l^r_-\ ,\cr\cr
\d\l^r_- \se -{1\over 2a} \c^\m \del_\m \phi^{+\,ir} \C_i \ve_+
+ 4ma \phi^{+\,ir} \C_i \ve_-\ , }
\la{sigma1}
\end{cases}
\w4
\hspace{-2cm}\mbox{Linearized Type 2:}&&\begin{cases}{
\d\phi^{-\,ir} \se {a\over 4m(1-2a^2)}\bar{\ve}_+\C^i\c^\m \del_\m
\l^r_+ + a\bar{\ve}_-\C^i\l_+^r\ , \cr\cr
\d\l^r_+ \se -2ma^{-1}(1-2a^2) \phi^{-\,ir}\C_i\ve_+\ .}
\la{sigma2}
\end{cases}
\eea

We now turn to the full transformation rules. In the case of the
Type 1 multiplet all the nonlinearities vanish except the
$A_{\m+}$ contribution to the special supersymmetry
transformation, and we find the local transformation rules

\be
\mbox{Full Type 1:}\quad\begin{cases}{
\d\phi^{+\,ir} \se a \eb\,\C^i \l^r_-\ ,\cr\cr
\d\l^r_- \se -{1\over 2a} \c^\m \del_\m \phi^{+\,ir} \C_i \ve
+ 4a \phi^{+\,ir} \C_i \eta\ , }
\la{sigma11}
\end{cases}
\ee

where $\eta$ is given by \eq{nota}. Note that the gauge field
$A_\m$ does not appear in the derivative of the scalars
$\phi^{ir}$ in the above formula because the chirality of $\ve$
projects it to its positive Hodge dual, which  is then absorbed
into the special supersymmetry parameter $\eta$. In other words,
the above formulae are $U(1)$ covariant modulo $\eta$
transformations.

In the Type 2 case the first nontrivial order the $D=3$ Dirac
equation now yields the following expression for the negative
chirality spinor:

\be \l^r_{(1)-} \se {1\over 4m(1-2a^2)} \c^\m {\Hat D}_\m \l_{+}^r
+{1\over 8ma^2}\C^3\c^\m A_{\m+}\l^r_{+}\ , \ee

where ${\Hat D}_\m \l$  is the supercovariant derivative defined
by

\be
{\Hat D}_\m \l^r = \nabla_\m \l^r +2ma^{-1}(1-2a^2) \f^{ir} \C_i \psi_\m\ .
\ee

The fact that matter fermions have the $U(1)$ charge $-1+{1\over
2a^2}$ has been used and $\nabla_\m$ is the ordinary Lorentz
covariant derivative.  Thus the full transformation rules for the
Type 2 multiplet reads:

\be
\mbox{Full Type 2:}\quad\begin{cases}{
\d\phi^{-\,ir} \se {a\over 4m(1-2a^2)}\bar{\ve}\C^i\c^\m {\Hat D}_\m
\l^r_+ + {a\over m}{\bar \eta}\C^i\l_+^r\ ,
\cr\cr
\d\l^r_+ \se -2ma^{-1}(1-2a^2)\phi^{-\,ir}\C_i\ve\ .} \la{sigma22}
\end{cases}
\ee

In summary, the transformation rules \eq{bt} and \eq{csg} are
those of $N=(2,0)$ conformal supergravity \cite{bns} consisting of
fields $(e_\m{}^a, \psi_{\m+}, A_{\m-})$ coupled to either one of
the following matter multiplets:

\begin{itemize}

\item{Type 1:} scalar multiplets consisting of scalar fields
$\phi^{ir}$ with Weyl weight $2a^2$ and negative chirality spinors
$\l^r_-$ with Weyl weight $\ft12+2a^2$.

\item{Type 2:} scalar multiplets consisting of scalar fields
$\phi^{ir}$ with Weyl weight $2-2a^2$ and positive chirality
spinors $\l^r_+$ with Weyl weight $\ft32-2a^2$.

\end{itemize}

The supersymmetry transformation rules \eq{csg}, \eq{sigma11} and
\eq{sigma22} close off-shell as follows

\bea
[\d_{\ve_1}, \d_{\ve_2}] &=& \d_\xi (\xi^\m) +
\d_{\L_L}(-\xi^\m \o_\m)+\d_\ve (-\xi^\m \psi_\m)\ ,
\label{c0}\w3
[\d_\eta,\d_\ve] &=& \d_{\L_D}(-2\eb\eta\,) +\d_{\L_L}(-2\eb\eta\,)
+\d_\L (4a^2\eb\C_3\eta\,)+\d_\eta(\ft12{\bar\eta}\psi^\m \c_\m\ve)\ ,
\la{c}
\eea

where $\xi^\m = \eb_1\c^\m\e_2$. The arguments on the right hand
side are the composite parameters of the relevant transformations.
Note the absence of the usual field dependent $U(1)$ gauge
transformation in the commutator of two supersymmetry
transformations. In obtaining \eqs{c0}{c} we have used

\be
\d \o_\m = \ft12 \eb\c^\l \psi_{\l\m}
+2{\bar\psi}_\n\c_\m\c^\n\eta\ ,
\ee

which directly follows from $\o_\m{}^{ab}=\o_\m \e^{ab}$ which is
determined from its algebraic equation of motion as

\be
\o_\m= e^{-1}\e^{\r\s} e_\m{}^a \del_\r e_{\s a}
+{\bar \psi}_\m\c^\n\psi_\n\ .
\la{o}
\ee

The result \eqs{c0}{c} is up to cubic fermion terms that may arise
through some of the composite $\ve$ and $\eta$-transformations,
since the transformations \eq{s2} were themselves up to that
order. In the case of Type 1 we have supercovariantized the
derivative of $\f^{+ir}$. In the case of Type 2, the
supercovariant derivative of $\l^r_+$ is already present, due to
the fact that the three-dimensional fermionic matter field
equation \eq{e4} is already supercovariant. However, in this case
the closure of two supersymmetries on the fermion requires $\d \f
\l$ type terms in $\d \l$ that are expected to arise in the
complete transformation rules.


\subsection{ Case 2\,:\ \ $\e=-1$ and $a^2=\ft12$ }


In this case, we recall the boundary behaviors of the matter
fields from \eq{bc2} and \eq{fex2} as follows:

\bea
\phi^{ir} & = & 2mr \left[ \phi_{(0)}^{+ir} + (2mr)^2
\f_{(2)}^{+ir} + \cdots \right] + 2mr \ln(2mr) \left[
\f_{(0)}^{-ir} + (2mr)^2 \f_{(2)}^{-ir} + \cdots \right], \nn\w2
\lambda^r & = & (2mr)^{1 \over 2} \left[ \lambda^r_{(0)+} + 2mr
\lambda^r_{(1)-} + \cdots \right] + (2mr)^{3 \over 2} \ln(2mr)
\left[ \l^r_{(0)-} + 2mr \l^r_{(1)+} + \cdots \right].
\label{matter}
\eea

As we shall show, the regular boundary fields
$(\f^{-ir}_{(0)},\l^r_{(0)+})$ form a supermultiplet of Type 1,
using the terminology introduced above, according to which
multiplets containing the regular scalars are called Type 1 and
those containing the irregular scalars are called Type 2. As for
the irregular fields $(\f^{+ir}_{(0)},\l^r_{(1)-})$, they will be
shown to form an extended multiplet, together with the fields of
the Type 1 multiplet.

Substituting \eq{matter} into the negative chirality component of
the $\lambda$ field equation we find that there are indeed no
conditions on $\l^r_{(0)+}$ and $\l^r_{(1)-}$ and that
$\l^r_{(0)-}$ is determined as follows:

\be
\l_{(0)-}^r
= {1 \over 2m} \c^\mu \hat D_\mu \l_{(0)+}^r,
\label{chiminus}
\ee

where

\be
\hat D_\mu \lambda_{(0)+}^r = \nabla_\m \lambda_{(0)+}^r +
\sqrt{2} m \f_{(0)}^{-ir}\Gamma_i \psi_{(0)\mu+} \ .
\ee

As we shall see later, $\hat D_\mu \lambda_{(0)+}^r$ is
supercovariant. Note that the U(1) charge $(-1+\ft{1}{2a^2})$
vanishes here since $a^2=\ft12$.

Let us now examine the boundary behaviour of the local symmetry
transformations, starting with the bosonic ones. The bosonic
transformations of the matter fields are the same as in the $0 <
a^2 < {1 \over 2}$ case, except for the dilatation which acts as
follows:

\bea
\hspace{-2cm}\mbox{Extended Type 1:}&&\begin{cases}{
\delta \f_{(0)}^{-ir}  =  - \Lambda_D \f_{(0)}^{-ir}\ ,
\cr\cr
\delta \lambda_{(0)+}^r = -{1 \over 2} \Lambda_D \lambda_{(0)+}^r\ ,
\cr\cr
\delta \phi_{(0)}^{+ir} =  - \Lambda_D \phi_{(0)}^{+ir} - \Lambda_D
\f_{(0)}^{-ir}\ ,
\cr\cr
\delta \lambda_{(1)-}^r =  - {3 \over 2} \Lambda_D \lambda_{(1)-}^r
- {1 \over 2m} \Lambda_D \c^\mu \hat D_\mu \l_{(0)+}^r \ .}
\la{dil}
\end{cases}
\eea

We see that due to the logarithmic terms in the expansion and the
fact that $\D_+=\D_-$ there is an admixture of the Type 1 fields
in the transformations of the irregular fields
$(\f^{+ir}_{(0)},\l^r_{(1)-})$. Thus, the full set of fields are
considered to form an extended Type 1 multiplet. This multiplet
structure will also emerge in the conformal supersymmetry
transformation rules.

We next turn to the boundary limits of the supersymmetry
transformations. Those of the supergravity multiplet take the
same form as given in \eq{csg}, with the replacement $A_{\m-} \ra
A'_{\m-}$ defined as (we suppress the $(0)$ labels in the
supergravity sector):

\be A'_{\mu -} = A_{\mu -} - {1 \over 4}
\bar\lambda_{(0)+}^r \gamma_\mu \Gamma^3 \lambda_{(0)+}^r\ .
\la{ap} \ee

To see this, we begin by noting that there is an additional log term
in the expansion of the Rarita-Schwinger field. By substituting
(\ref{matter}) into the $r\mu$ component of the Rarita-Schwinger
field equation we find

\bea
\psi_{\mu+}(r, \vec x) & = & (2mr)^{-{1 \over 2}}
\psi_{\mu+} + \cdots,
\nn\w2
\psi_{\mu-}(r, \vec x) & = &
(2mr)^{1 \over 2} \psi_{\mu-} + {1 \over \sqrt{2}} (2mr)^{1 \over
2} \ln(2mr) \gamma_\mu \f_{(0)}^{-ir}\Gamma_i \lambda_{(0)+}^r  +
\cdots.
\eea

The $\mu\nu$ component of the Rarita-Schwinger field equation
gives

\be
{m\over 2}\c_\m\c^\n\psi_{\n-} = \ft14 \c^\n\psi_{\m\n}
+{1 \over 4} \C^3\c_\m\psi^\n A_{\n +}
+ {1 \over \sqrt{2}} m \c_\m
\left( \f_{(0)}^{+ir} + {1 \over 2} \f_{(0)}^{-ir} \right)
\C_i \l^r_{(0)+}\ ,
\ee

where $\psi_{\m\n}=D_\m\psi_\n -D_\n\psi_\m$ and $D_\m \ve=(\nabla_\m
-A_\m \C_3) \ve $. Using this the fermionic transformation of
$A_{\mu-}$ becomes

\be
\d A_{\mu-} = {1 \over 2} \bar\ve\C^3\c^\nu \psi_{\mu\nu}
+ \bar\eta \C^3 \c^\nu \c_\mu \psi_\nu
-{1 \over \sqrt{2}} m \bar\ve\c_\m\C_i\C^3 \l_{(0)+}^r \f_{(0)}^{-ir}\ .
\ee

The matter dependence can be removed by the redefinition \eq{ap}.
The field $A'_{\m -}$ then transforms as in \eq{csg} upon the use
of $\l_{(0)+}^r$ given below. Altogether, we find the conformal
supergravity multiplet transformations \eq{csg}, with $a^2=\ft12$.

We next study the supersymmetry transformations of the matter
fields near the boundary. We find after rescaling $\f^{\pm} \ra
\ft1{\sqrt 2}\f^{\pm}$ (and dropping the $(0)$'s for notational
simplicity)

\be
\mbox{Extended Type 1:} \quad
\mx{\{}{lcl}{
\delta\f^{-ir} & = & {1 \over 2m} \bar\ve
\Gamma^i\gamma^\mu \hat D_\mu \lambda_{+}^r\ ,
\nn\w3
\delta \l_+^r & = & -m \f^{-ir}\Gamma_i \ve \ ,
\nn\w3
\delta \phi^{+ir} & = &  \bar\ve\Gamma^i
\lambda_-^{'r} + {1 \over m} \bar\eta \Gamma^i \lambda_+^r\ ,
\nn\w3
\delta \lambda_-^{'r} & = & - {1 \over 2} \gamma^\mu
\partial_\mu \phi^{+ir} \Gamma_i \ve +
2\Gamma_i \eta\,( \phi^{+ir}+ {1\over 2} \f^{-ir}) \ ,}{.}
\la{et1}
\ee

where we have used the field equation \eq{chiminus} and made the
field redefinition

\be \lambda_{-}^{'r} = \lambda_{(1)-}^r - {1 \over 4m} \gamma^\mu
A_{\mu+} \Gamma^3 \lambda_{(0)+}^r. \ee

We observe that $(\f^{-ir},\l_+^r)$ form a Type 1 submultiplet,
whose fields are inert under the conformal supersymmetry
transformations, and whose supersymmetry transformations can be
obtained by rescaling the fields of the Type 2 multiplet
\eq{sigma22} as $\f^{-ir}\ra (1-2a^2)^{-1}\f^{-ir}$ and $\l_+\ra
2{\sqrt 2}\l_+$, and taking the limit $a^2\ra \ft12 $.
Correspondingly, the superalgebra closes as in \eqs{c0}{c} by
setting $a^2=\ft12$. The fields $(\f^{+ir},\l_-^{'r})$, on the
other hand, transform into each other under the ordinary
supersymmetry transformations, but transform into $(\f^{-ir}$,
$\lambda_+^r)$ under the conformal supersymmetry transformations.
(This is similar to the situation of the dilatation transformation
laws \eq{dil}}. Therefore, we view the enlarged set of fields
$(\f^{+ir}, \f^{-ir}, \l_-^{'r}, \l_+^r)$ as forming a conformal
supermultiplet, with closure given by \eqs{c0}{c} for $a^2=\ft12$
(with dilatation transformation \eq{dil}). This extended Type 1
multiplet can be truncated consistently to an ordinary Type 1
multiplet by setting the Type 2 submultiplet equal to zero, but
the reverse is not consistent.


\subsection{ Case 3\,:\ \ $a^2=0$  }


In this case the boundary behavior of the matter fields is given
by

\bea
\phi & = & \left[
\phi_{(0)}^+ + (2mr)^2 \phi_{(2)}^+ + \cdots \right]
+ (2mr)^2 \ln(2mr) \left[
\phi_{(0)}^- + (2mr)^2 \phi_{(2)}^- + \cdots \right],
\nn\w2
\lambda & = & (2mr)^{1 \over 2} \left[ \lambda_{(0)-}
+ 2mr \lambda_{(1)+} + (2mr)^2 \lambda_{(2)-} + \cdots \right]
\nn\w2
&& + (2mr)^{3 \over 2} \ln(2mr) \left[
\lambda_{(0)+} + 2mr \lambda_{(1)-} + \cdots \right]\ .
\label{matter2}
\eea

The supersymmetric variations of the zweibein and gravitino are as
in \eq{csg} while the transformation of $A_{\mu-}$ has a subtlety
due to the fact that  there is an additional log term in the
expansion of the Rarita-Schwinger field. Taking this into account, from
the Rarita-Schwinger field equation we find

\bea
\psi_{\mu+}(r, \vec x)
& = & (2mr)^{-{1 \over 2}} \psi_{\mu+} + \cdots,
\nn\w2
\psi_{\mu-}(r, \vec x)
& = & (2mr)^{1 \over 2} \psi_{(0)\mu-}
+ {1 \over 4m} (2mr)^{1 \over 2} \ln(2mr) \Gamma_i
\gamma^\nu \gamma_\mu \lambda_{(0)-}^r
\partial_\nu \phi_{(0)}^{+ir} + \cdots\ ,\\
{m\over 2}\c_\m\c^\n\psi_{(0)\n-} &=& \ft14
\c^\n\psi_{\m\n+} +{1 \over 8} \C^3 \c_\m \psi_{+}^\n
A_{\n+} - {1 \over 8} \Gamma_i \gamma_\mu \gamma^\nu \l^r_{(0)-}
\partial_\nu \phi_{(0)}^{+ir}\ ,
\eea

where $\psi_{\m\n}=D_\m\psi_\n -D_\n\psi_\m$ with the U(1)
connection term in $D_\m\psi$ shifted as in \eq{d3}. Using this the
fermionic transformation of $A_{\mu-}$ becomes

\be
\delta A_{\mu-} = \bar\varepsilon \Gamma^3
\gamma^\nu \psi_{(0)\mu\nu+}
+ 2 \bar\eta \Gamma^3 \gamma^\nu \gamma_\mu \psi_{(0)\nu+}
- {1 \over 2} \bar\varepsilon \Gamma^3 \Gamma_i \gamma_\mu
\gamma^\nu \lambda_{(0)-}^r \partial_\nu \phi_{(0)}^{+ir}\ ,
\ee

where we have used \eq{da}. To cancel the matter contributions, we
first need to examine the transformations of the matter fields.
>From the matter field equations and the boundary conditions stated
earlier for the case at hand, we find that the independent fields
at the boundary are $(\phi_{(0)}^{+ir}, \lambda_{(0)-}^r,
\phi_{(2)}^{+ir}, \lambda_{(1)+}^r)$. Under dilatations, these
fields are found to transform as

\bea
\delta \phi_{(0)}^{+ir} & = & 0,
\nn\w2
\delta\lambda_{(0)-}^r & = & - {1 \over 2} \Lambda_D \lambda_{(0)-}^r,
\nn\w2
\delta \phi_{(2)}^{+ir} & = & - 2 \Lambda_D
\phi_{(2)}^{+ir} - \Lambda_D \phi_{(0)}^{-ir},
\nn\w2
\delta \lambda_{(1)+}^r & = & - {3 \over 2} \Lambda_D \lambda_{(1)+}^r -
\Lambda_D \lambda_{(0)+}^r\ ,
\label{dilatation}
\eea

where, again, a mixing of the type observed earlier in the case of
$a^2=\ft12$ arises here. The fermionic transformation of the
matter fields also exhibit this kind of mixing:

\bea
\delta \phi_{(0)}^{+ir} & = & \bar\varepsilon_{(0)+}
\Gamma^i \lambda_{(0)-}^r\ ,
\nn\w2
\delta \lambda_{(0)-}^r & = &
- {1 \over 2} \Gamma_i \gamma^\mu \varepsilon_{(0)+} \partial_\mu
\phi_{(0)}^{+ir}\ ,
\nn\w2
\delta \phi_{(2)}^{+ir} & = &
\bar\varepsilon_{(0)+} \Gamma^i \lambda_{(2)-}^r +
\bar\varepsilon_{(0)-} \Gamma^i \lambda_{(1)+}^r\ ,
\nn\w2
\delta\lambda_{(1)+}^r & = & - {1 \over 2} \Gamma_i \gamma^\mu
\varepsilon_{(0)-}
\partial_\mu \phi_{(0)}^{+ir}
- 2m \Gamma_i \varepsilon_{(0)+} \left( \phi_{(2)}^{+ir} +
{1 \over 2} \phi_{(0)}^{-ir} \right) \ ,
\la{susyc}
\eea

where
$\phi_{(0)}^{-ir}=-\ft1{8m^2}\nabla^\m\del_\m\phi_{(0)}^{+ir}$ and
$\lambda_{(2)-}^r$ is a more complicated function of the
independent fields.

Armed with this result, we first redefine $A_\mu$ as in \eq{defa}
so that the newly defined field $A'_{\mu-}$ transforms precisely
as in \eq{csg}, that is without any matter contributions, as
expected from an off-shell conformal supergravity multiplet. As
for the interpretation of the matter multiplet transformations,
surprisingly enough, the story is somewhat more complicated. While
the $\ve_{(0)-}$ parameter can be redefined into the special
supersymmetry parameter $\eta$ as in \eq{nota}, a close
examination of the transformation rules $\d\phi_{(2)}^{+ir}$ and
$\d\lambda_{(1)+}^r$ shows that the dependence of the result on
$A_{\mu+}$ and the special supersymmetry gauge field, which is an
appropriately redefined $\psi_{\mu-}$, cannot be removed unless
certain equations of motion are imposed. However, these gauge
fields are determined in terms of the independent fields as
nonlocal expressions. Therefore, in order to realize the conformal
supersymmetry on the boundary in a local fashion, we need to
remove the dependence on these dependent gauge fields. This can be
achieved by setting

\be
\left( g^{\m\n}- \frac{\e^{\m\n}}{\sqrt{-g}}\right)\,
{\hat D}_\n\phi_{(0)}^{+ir}=0\ , \quad\quad \c^\m {\hat D}_\m \l_{(0)-}^r=0\ ,
\la{os}
\ee

where the supercovariant derivatives are defined in a standard way
in accordance with the supersymmetry variations \eq{susyc}, and
${\hat D}_\n\l_{(0)-}^r$ contains the shifted field $A'_{\m -}$.
These field equations transform into each other under the
supersymmetry variations \eq{susyc}, as they should.

Imposing the on-shell conditions \eq{os}, and recalling that the
derivatives on \eq{susyc} need to be supercovariantized when
considering the higher order fermion terms, we find that the
$\ve_{(0)-}$ term in the last equation in \eq{susyc} drops out,
and that the boundary evaluation of the $D=3$ matter field
equations imply that

\bea
\phi_{(0)}^{-ir} &=&0\ ,\qquad \l_{(0)+}^r = 0\ ,
\la{os1}
\w2
\l_{(2)-}^r &=& {1\over 4m} \c^\mu \left( \hat D_\mu \l_{(1)+}^r +
{1 \over 2} A_{\mu+} \C^3 \l_{(1)+}^r \right)\ ,
\la{os2}
\eea

with the supercovariant derivatives defined in standard way.
Substituting these results in \eq{susyc}, the $A_{\mu+}$ dependent
term can be absorbed into redefinition of $\ve_{(0)-}$ to yield
the special supersymmetry parameter $\eta$, and all in all, the
supersymmetry transformation rules \eq{susyc} for the independent
field disentangle into those of two separate multiplet of fields
as follows (dropping the $(0)$ and chirality labels on the
conformal supergravity fields and parameters):

\bea
\hspace{-2cm}\mbox{Type 1:}&&\begin{cases}{
\d\phi_{(0)}^{+ir} =  \bar\ve \C^i \l_{(0)-}^r\ ,
\cr\cr
\d\l_{(0)-}^r  =  -\ft12 \C_i \c^\mu \ve \del_\mu
\phi_{(0)}^{+ir}\ ,}
\la{zero1}
\end{cases}
\w4 \hspace{-2cm}\mbox{Type 2:}&&\begin{cases}{
\d\phi_{(2)}^{+ir} =
{1 \over 4m} \bar\ve \C^i \c^\mu \hat D_\mu \lambda_{(1)+}^r
 + {1\over m} \bar\eta \C^i \l_{(1)+}^r\ ,\cr\cr
\d\l_{(1)+}^r  =  - 2m \C_i \ve \phi_{(2)}^{+ir}\ .}
\la{susyd}
\end{cases}
\eea

Again, we have used the terminology of Type 1 and Type 2,
according to whether the multiplet contains regular or irregular
scalar fields.  The result for Type 1 agrees with that of
\cite{bns}, where (2,0) conformal supergravity and its coupling to
a sigma model in $D=2$ is constructed. All the fields occurring in
both multiplets above now have definite Weyl weights since the
mixings in the dilatation transformations \eq{dilatation}
disappear upon the use of \eqs{os1}{os2}.

The algebra \eqs{zero1}{susyd} closes as in \eq{c}. In
interpreting the composite U(1) transformation, the composite
parameter must be rescaled by $a^2$, since $A_\mu$ has been
rescaled by $a^2$, prior to taking the limit $a^2 \ra 0$. The
closure of the algebra can be seen from the fact that the Type 1
and Type 2 multiplets here can be obtained from the Type 1 and
Type 2 multiplets found Section 5.1 for $0 <a^2<\ft12$, by first
rescaling the scalar fields as $\phi \ra a\phi$ and then taking
$a^2\ra 0$.


\section{ Conclusions }


In this paper we have analyzed the behaviour of (2,0) gauged
supergravity coupled to matter in $D=3$ near the boundary of AdS.
We have exhibited the role of  the bulk supergravity and matter
field equations in determining the realization of conformal
supersymmetry on the  boundary of AdS. We have found that various
types of matter multiplets emerge at the boundary in addition to a
universal (2,0) conformal supergravity multiplet These multiplets
involve fields whose conformal dimensions depend on the radius of
the K\"ahlerian sigma model coset space and on the gravitational
coupling constant (set equal to 1 in most of the paper). The
nature of the boundary conformal multiplets found depends crucially
on the ratio of these constants. Interestingly, the local
supersymmetry of the $D=3$ theory does not fix this ratio nor the
sign of the sigma model curvature constant, though the most
interesting boundary conditions turn out to be possible for
noncompact sigma model coset space whose curvature scalar is
restricted to lie in a finite range in units of the $D=3$ Planck
length, as discussed in Section 5. In the case of flat sigma model
manifold, we find a connection between the model of \cite{mc} and
that of Izquierdo and Townsend \cite{it}.

We have seen that there are  several subtleties in choosing the
boundary conditions for the matter fields. In particular, we find
that both regular and irregular boundary conditions can be imposed
on the matter fields as a consequence of the fact that scalar
fields with  sufficiently negative mass-squared can be associated
with CFT operators of two possible dimensions on the boundary. In
fact, this phenomenon has already been observed in \cite{kw} in
the context of $AdS_5 \times T^{1,1}$ compactification of Type IIB
string theory. Here, we provide another example of this
phenomenon, and we find the resulting CFT supergravity plus matter
symmetry transformations. Somewhat surprisingly, we also find an
interesting conformal supermultiplet structure on the boundary
that involve fields which do not have definite Weyl weights but
rather mix with other fields of the multiplet under dilatations.
In this novel multiplet the superconformal symmetry is also
realized in an unconventional fashion.

In the case of irregular boundary conditions the analysis had to
be restricted for certain values of the sigma model radius,
referred to as Case 1 in Section 5, such that the effects of the
nonlinear contributions from the regular fields to the
transformations of the irregular fields were omitted. The
inclusion of both regular and irregular fields is necessary for
the interpretations of the AdS/CFT correspondence. The study of
these effects is intimately connected with the identification of
the boundary conformal field theory, which lies beyond the scope
of this paper.

We conclude by commenting on some of the interesting open
problems. Firstly, it is clearly desirable to find an M-theoretic
origin of the  model studied here. The structure of the conformal
supermultiplets that we have found on the boundary provide
information on a class of operators which the boundary CFT must
contain but do not provide the full data required to specify
uniquely the the CFT in question. It is conceivable that an
M-theoretic origin of the model exists only for a certain critical
value of the sigma model curvature constant. At any rate, many of
the features encountered in the analysis of the $(2,0), AdS_3$
supergravity plus matter system studied here are likely to arise
in the $(4,4), AdS_3$ supergravity plus matter system which arises
in the $AdS_3 \times S^3$ compactification of $(2,0), D=6$
supergravity coupled to tensor multiplets, whose embedding in
M-theory is known. We hope that the results presented here may
give a flavor of what to expect in that case. Indeed, these
results may also prove useful in analysing higher dimensional AdS
supergravity plus matter systems as well.

It would also be interesting to extend the above analysis to a
generalized setup in which the boundary conditions are imposed on
a surface which is a finite distance away from the AdS boundary.
This is expected to provide an understanding of how a supergravity
plus matter system can be localized on a brane worldvolume in a
Randall-Sundrum like scenario. This leads to normalizable bulk
modes which correspond to fluctuating boundary modes. The boundary
CFT, which is dual to the matter coupled supergravity in the bulk,
should therefore be supplemented by an off-shell Lagrangian for
the (fluctuating) boundary supergravity/matter modes.
Thus, the total dynamics is that of the boundary CFT plus the
localized, matter coupled bulk supergravity. In this context
the unexpected result of Section 5.3, in the form of the on-shell
constraint given in \eq{os}, indicates that ordinary off-shell as
well as chiral two-dimensional matter systems may be localized on
the brane. We finally remark that we expect the proper vacuum for
this setup to be the black string (domain wall) solution of
\cite{mc}, rather than the anti-de Sitter vacuum. It would
therefore be interesting to extend the bulk theory by the
inclusion of two-form potentials and to give the supersymmetric
coupling of these black strings to the bulk supergravity
\cite{bkvp}

\bigskip \bigskip


 \noindent{\large \bf Acknowledgments}


\medskip

We thank Kostas Skenderis for valuable discussions. The work of
E.S. has been supported in part by NSF Grant PHY-0070964, the
work of Y.T. in part by the Grant-in-Aid for Scientific Research
on Priority Area 707 ``Supersymmetry and Unified Theory of
Elementary Particles'', Japan Ministry of Education, and the work
of P.S. in full by Stichting FOM, Netherlands.

\pagebreak


\end{document}